\documentclass[journal=apchd5,manuscript=article]{achemso}

\usepackage[version=3]{mhchem} % Formula subscripts using \ce{}
\usepackage[T1]{fontenc}       % Use modern font encodings
\usepackage{microtype} %Makes kerning and spacing even better than usual in Latex.
\usepackage{color}

\author{Martin Th\"{a}mer}
\affiliation[FHI]{Fritz Haber Institute of the Max Planck Society, 4-6 Faradayweg, 14195 Berlin, Germany}
\email{thaemer@fhi-berlin.mpg.de}
\author{R.\ Kramer Campen}
\affiliation[FHI]{Fritz Haber Institute of the Max Planck Society, 4-6 Faradayweg, 14195 Berlin, Germany}
\author{Martin Wolf}
\affiliation[FHI]{Fritz Haber Institute of the Max Planck Society, 4-6 Faradayweg, 14195 Berlin, Germany}

\title[Phase Certainty and High Signal to Noise]
  {Detecting Weak Signals from Interfaces by High Accuracy Phase-Resolved SFG Spectroscopy}

\abbreviations{IR,NMR,UV}
\begin{document}

%\begin{tocentry}

%Some journals require a graphical entry for the Table of Contents.
%This should be laid out ``print ready'' so that the sizing of the

%Inside the \texttt{tocentry} environment, the font used is Helvetica
%8\,pt, as required by \emph{Journal of the American Chemical

%The surrounding frame is 9\,cm by 3.5\,cm, which is the maximum
%permitted for  \emph{Journal of the American Chemical Society}
%graphical table of content entries. The box will not resize if the
%content is too big: instead it will overflow the edge of the box.

%This box and the associated title will always be printed on a
%separate page at the end of the document.

%\end{tocentry}

\begin{abstract}
Much work over the last 25 years has demonstrated that the interface-specific, all-optical technique, vibrational sum frequency generation (v-SFG) spectroscopy, is often uniquely capable of characterizing the structure and dynamics of interfacial species. The desired information in such a measurement is the complex second order susceptibility which gives rise to the nonlinear response from interfacial molecules. The ability to detect molecular species yielding only small contributions to the susceptibility is meanwhile limited by the precision by which the spectral phase and amplitude can be determined. In this study we describe a new spectrometer design that offers unprecedented phase and amplitude accuracy while significantly improving the sensitivity of the technique. Combining a full collinear beam geometry with a technique enabling the simultaneous measurement of the complex sample and reference spectrum, uncertainties in the reference phase and amplitude are shown to be greatly reduced. Furthermore, we show that using balanced detection, the signal to noise ratio can be increased by one order of magnitude. The capabilities of the spectrometer are demonstrated by the isolation of a small isotropic surface signal from the bulk dominated nonlinear optical response of z-cut quartz. The achieved precision of our spectrometer enables measurements not currently feasible in v-SFG spectroscopy.
\end{abstract}

\section{Introduction}
The macroscopic properties of a wide variety of biophysical, environmental and engineered systems hinge on the behavior of molecules at solid or liquid interfaces. Characterizing the structure and dynamics of these interfacial species under their native environmental conditions is a formidable experimental challenge. Because the interfaces in such systems are typically buried under some material (\textit{e.g.}\ air, aqueous solutions, or organic tissue) the use of atoms or electrons as probes is usually precluded. In these cases nondestructive, all-optical spectroscopies are natural candidates to gain such insight. However, the lack of interfacial sensitivity of linear approaches often makes it difficult to distinguish the spectral response of molecules at interfaces from the similar spectral response of a much larger numbers of molecules in the adjoining bulk phases. Much work in the last 25 years has demonstrated that the laser-based, nonlinear optical technique, vibrational sum frequency generation (v-SFG) spectroscopy enables probing the spectral response of molecules with interfacial specificity \cite{zhu87,she89,bai95,lam05,zha94,ric02,hun87,sto09,che02,dav10}.

To perform a v-SFG measurement infrared, $\text{E}_{\text{IR}}({{\omega }_{\text{IR}}})$, and visible, $\text{E}_{\text{vis}}({{\omega }_{\text{vis}}})$, laser pulses are overlapped spatially and temporally at an interface and the field emitted at the sum of the frequencies of the two incident beams detected $\text{E}_{\text{SFG}}({{\omega}_{\text{SFG}}})$. 
%This sum frequency generation process conserves energy, $\omega_{\text{sfg}} = \omega_{\text{vis}} + \omega_{\text{ir}}$ and wave vector, $n_{\text{sfg}}\vec{k}_{\text{sfg}} = n_{\text{vis}}\vec{k}_{\text{vis}} + n_{\text{ir}}\vec{k}_{\text{ir}}$ (in which $n_{i}$ indicates the refractive index of the medium through which the $i$ beam propagates and $\vec{k}_{i}$ is its wave vector). The latter condition means that the angle of propagation of the emitted sfg is given by the frequencies and angles of the incident beams and the, linear, refractive indices of all relevant media. 
%The relationship of the emitted sum frequency field to the incident fields is,
%
\begin{equation}\label{e:basic}
{{\text{E}}_{\text{SFG}}} \propto {{\chi }^{(2)}}({{\omega}_{\text{SFG}}}={{\omega}_{\text{IR}}}+{{\omega}_{\text{vis}}}):\text{E}_{\text{vis}}\text{E}_{\text{IR}}
%{{E}_{\text{sfg}}}({{\omega}_{\text{sfg}}})\propto {{\chi }^{(2)}}({{\omega}_{\text{sfg}}}={{\omega}_{\text{ir}}}+{{\omega}_{\text{vis}}}):E_{\text{vis}}({{\omega}_{\text{vis}}})E_{\text{ir}}({{\omega}_{\text{ir}}})
\end{equation}
The spectral phase and amplitude of the generated SFG signal is a function of the phases and amplitudes of the interacting light pulses and the complex second order susceptibility ${{\chi }^{(2)}}$. Latter contains the desired spectroscopic information.\cite{she03} Its imaginary part describes resonances in the sample and carries similar information as a typical absorption spectrum. The interface sensitivity originates meanwhile from the symmetry properties of ${{\chi }^{(2)}}$. The sign of its imaginary part is related to the orientation of the oscillating dipole\cite{she13,she06}. By flipping orientation of the dipole by 180 deg. the SFG signal changes its phase by 180 deg. resulting in a sign flip of the second order susceptibility. This leads to the cancellation of the SFG contributions in presence of centro-symmetry (under electric dipole approximation). Consequently, if v-SFG is applied to a sample consisting of two different centro-symmetric or amorphous media the generated SFG signal purely originates from the interface regions where the symmetry is broken. Applying v-SFG spectroscopy to such samples consequently yields interfacial specificity combined with information on the orientation of the corresponding species. 

v-SFG spectroscopy at interfaces is usually performed such that there are no sample resonances at either the visible and the SFG frequencies. The second order susceptibility is then typically dominated by vibrational resonances yielding a vibrational fingerprint of the interfacial species. By analyzing its imaginary part deep insight into the molecular structure of an interface can be gained. The characteristic positions of resonance peaks allow in principle for the identification and characterization of interfacial molecular species and their amplitudes are related to the corresponding populations while the sign of the peaks reveals their orientation. For most samples, however, the measured $\chi^{(2)}$ is composed of a linear superposition of, possibly multiple, resonant and nonresonant contributions all of which are complex. Those individual contributions can largely differ in amplitude, phase, and symmetry which can make the resulting spectrum difficult to interpret. For its decomposition it is first essential to measure the complex $\chi^{(2)}$ and not its square modulus ${{\left| {{\chi }^{(2)}} \right|}^{2}}$ as it is the case in most common SFG spectrometers \cite{lam05,tia14,ric98,lag10}. Such homodyned techniques measure the intensity of the generated SFG signal without phase resolution which gives rise to undesired interference cross terms between the different contributions of the second order susceptibility. On the other hand, employing heterodyned techniques where the generated SFG field is interfered with a reference SFG signal (local oscillator) the phase information is preserved and the complex susceptibility can be determined \cite{she13,ji07,che10,nih09}. However, the isolation of the optical response from a specific molecular species can still be very challenging \textit{e.g.}\ if its contribution is not the dominating component in the complex spectrum of  $\chi^{(2)}$; in fact it is often the molecular species with low interface populations which are of particular interest (\textit{e.g.}\ protons at interfaces). The signal is then typically buried under much more intense spectral features and appears as a slight modification of the spectral phase and amplitude of the overall susceptibility. It is our aim to develop a phase sensitive SFG-spectrometer that is capable of resolving such small signals to enable the investigation of molecular species which only sparsely populate the interface.

A possible way to clearly identify a small spectroscopic signature of a particular molecular species in a vibrational spectrum is by altering its spectral response in one or a set of reference spectra. The reference could be a sample where the species is simply absent (or present with a different concentration) or where its vibrational resonances spectrally are shifted (\textit{e.g.}\ by isotope labeling). Another possibility is the distinction of the different contributions by different symmetry (changing angles and polarizations of the interacting laser pulses). All these techniques have in common that one needs to resolve a small change in spectral phase and amplitude in the measured overall ${{\chi }^{(2)}}$ between two or multiple acquired data sets. The different acquisitions might thereby involve the physical exchange of samples and/or subsequent scans of a sample under modified experimental conditions. The \textit{sensitivity}\ of the spectroscopy to detect the desired species is then given by the accuracy at which phase and amplitude of $\chi^{(2)}$ can be determined in the different acquisitions in combination with the signal to noise ratio achieved in the resulting spectra. 

Obtaining a high level of accuracy is, however, a tremendous experimental challenge as the relative phases of laser pulses tend to drift and also their intensities and spectra typically show significant changes with time. These temporal changes directly lead to phase and amplitude uncertainties between subsequent measurements and diminish their comparability. An elimination of such drifts by active or passive stabilization of the SFG spectrometer is rather unpractical because it involves extensive technical measures. It should be noted that very small changes of relative beam paths (by a few tens of nanometer) already lead to dramatic phase shifts at visible (SFG) frequencies. Another source of inaccuracy is related to the non-collinear beam geometry that is (for practical reasons) commonly implemented in heterodyned SFG spectrometers. It makes the phase and amplitude of the generated SFG signal sensitive to the sample position. As a consequence it is very challenging to maintain phase and amplitude accuracy upon exchange of samples \cite{nih13}. Furthermore, it makes it extremely difficult to obtain reliable data from liquid samples where the position of the interface may constantly change due to evaporation. Finally, the data recorded in heterodyned SFG spectroscopy typically contain considerable amount of noise. This has mainly two reasons. On the one hand SFG signals generated at interfaces are in general very weak and therefore difficult to detect, on the other hand compared to spectroscopic techniques employing incoherent or continuous wave light sources the SFG signals tend to show relatively large intensity fluctuations. These fluctuations result from the fact that the heterodyned SFG signal is generated by the nonlinear interaction of three ultrashort laser pulses with the sample (infrared pulse, visible upconversion pulse, local oscillator pulse). This process amplifies any types fluctuations of the initial pulses which can hardly be removed. Additional noise can originate from phase jitter in the interference between the local oscillator and the SFG signal. Overall, these technical limitations clearly reduce the ability of common phase sensitive SFG-spectrometers to resolve small spectral features which typically restricts v-SFG studies to the investigation of molecular species that yield large, dominating SFG signals.

In this study we describe a newly developed phase sensitive, time domain v-SFG spectrometer that addresses all of the technical challenges mentioned above. By combining a full collinear beam geometry with a method for simultaneous referencing we achieve unpreceded accuracy in phase and amplitude between sample and reference measurements. Moreover, employing the technique of balanced detection in combination with a special data treatment we very efficiently reduce noise. These improvements represent a significant technical advancement which will allow for phase resolved v-SFG studies of $\chi^{(2)}$ components which are too small to be detected via more conventional approaches.
 
\section{(Collinear) Time Domain SFG Spectrometer}
To acquire phase resolved v-SFG spectra we chose a rather unconventional heterodyned, full time domain approach because it offers considerable advantages compared to the more commonly used frequency domain techniques as discussed in reference  \cite{laa11}. 
\begin{figure}[htbp]
	\begin{center}
			\includegraphics[width=0.3\textwidth]{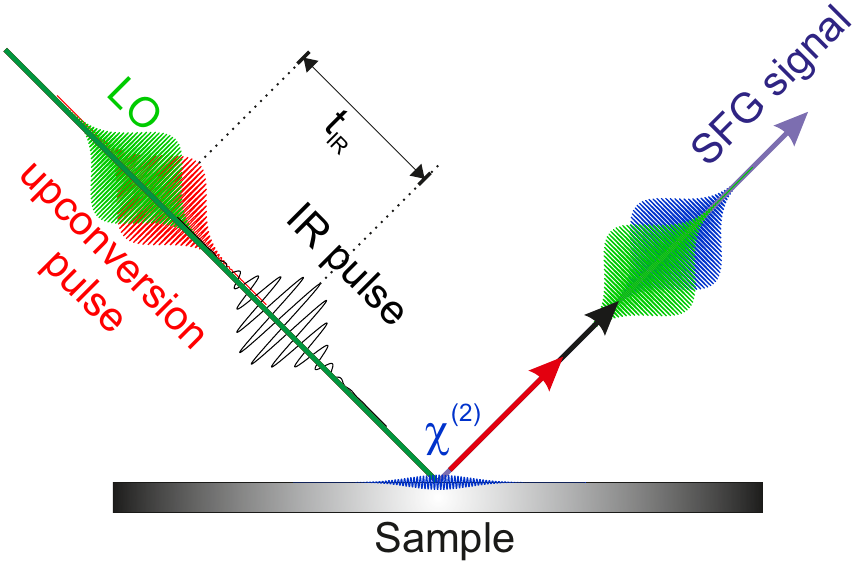}
		\caption{Schematic representation of time domain heterodyned SFG spectroscopy. The interaction of the infrared (black) and the upconversion pulse (red) with the sample generates an SFG signal (blue) which interferes with the local oscillator (green). Scanning the time delay t$_{\text{IR}}$ modulates the relative phases between the SFG signal and LO which produces the desired interferogram.} 
		\label{f:tds}
	\end{center}
\end{figure}

The heterodyned v-SFG signal is generated by the interaction of ultrashort/broadband infrared, upconversion and local oscillator pulses (see Figure \ref{f:tds}). The LO is produced by combining the upconversion pulse with a fraction of the infrared in a nonlinear crystal securing a well-defined phase relationship between all three pulses. A variable time delay $t_{IR}$ is introduced between the infrared and the upconversion pulses while the timing between the upconversion and LO pulses is fixed at nearly zero delay. The interaction of infrared and upconversion pulses with the sample generates a sum frequency signal that subsequently interferes with the LO. By scanning $t_{IR}$ the relative phase between the SFG signal and the LO is modulated at infrared frequencies producing an interferogram. Fourier transformation of the interferogram then yields the IR frequency resolved complex second order spectrum.  

The generated SFG field $\text{E}_{\text{SFG}}$ can mathematically be expressed as the convolution of the interacting light fields $\text{E}_{\text{vis}}$ and $\text{E}_{\text{IR}}$ with the second order response function $R^{\text{(2)}}$ of the sample (where $t_{1}$ is the time elapsed following the IR interaction, $t_{2}$ the time elapsed following the vis, $t_{\text{IR}}$ the delay between the IR and vis pulses and $t$ is time).
\begin{equation}\label{e:sig1}
\begin{aligned}
	\text{E}_{\text{SFG}}(t,t_{\text{IR}}) \propto \int_{0}^{\infty} dt_{2}\int_{0}^{\infty}dt_{1}R^{\text{(2)}}(t_{2},t_{1})\text{E}_{\text{vis}}(t-t_{2}+t_{\text{IR}})\\
\text{E}_{\text{IR}}(t-t_{2}-t_{1})
\end{aligned}
\end{equation}
Assuming no sample resonances at either, the upconversion- or the SFG frequency, $R^{\text{(2)}}$ can be split into the product of a resonant interaction with the infrared field $R^{\text{(2)}\dag}$ and a nonresonant interaction with the visible field. The response function of a nonresonant interaction can meanwhile be approximated by a delta function $\delta$.
\begin{equation}\label{e:R}
R^{\text{(2)}}(t_{2},t_{1}) \approx R^{\text{(2)}\dag}(t_{1})\delta(t_{2})
\end{equation}
Implementing this approximation into equation \ref{e:sig1} yields,
\begin{equation}\label{e:sig2}
	\text{E}_{\text{SFG}}(t,t_{\text{IR}}) \propto  \text{E}_{\text{vis}}(t+t_{\text{IR}}) \left[R^{\text{(2)}\dag}\otimes\text{E}_{\text{IR}}(t) \right]
\end{equation} 
with the convolution operator $\otimes$.
The measured heterodyned signal intensity $\text{I}_{\text{het}}$ is subsequently given by
\begin{equation}\label{e:het1}
	\text{I}_{\text{het}}(t_{\text{IR}}) \propto \int_{-\infty}^{\infty}dt \left(\text{E}_{\text{LO}}(t+t_{\text{IR}})+ \text{E}_{\text{SFG}}(t,t_{\text{IR}}) \right)^{2}
\end{equation}
where $\text{E}_{\text{LO}}$ denotes the local oscillator field. Filtering out all contributions \emph{except} the interference term (by balanced detection, see next section) reduces equation \ref{e:het1} to
\begin{equation}\label{e:het2}
	\text{I}_{\text{het,bal}}(t_{\text{IR}}) \propto \int^{\infty}_{-\infty}dt \left( \text{E}_{\text{LO}}(t+t_{\text{IR}})\text{E}_{\text{SFG}}(t,t_{\text{IR}})    \right)
\end{equation}
Combining equations \ref{e:het2} and \ref{e:sig2} leads to the following expression,
\begin{equation}\label{e:het3}
\begin{aligned}
	\text{I}_{\text{het,bal}}(t_{\text{IR}}) \propto \int^{\infty}_{-\infty}dt \left(R^{\text{(2)}\dag}\otimes\text{E}_{\text{IR}}(t)\left[\text{E}_{\text{vis}}(t+t_{\text{IR}})\text{E}_{\text{LO}}(t+t_{\text{IR}})\right]\right) \\
= R^{\text{(2)}\dag}\otimes \text{E}_{\text{IR}}(t)\circ \left[\text{E}_{\text{vis}}(t)\text{E}_{\text{LO}}(t)\right]
\end{aligned}
\end{equation}
where $\circ$ is the correlation operator. After Fourier transformation of equation \ref{e:het3} and under application of the convolution theorem we obtain
\begin{equation}\label{e:fhet1}
	\mathcal{F}\left(\text{I}_{\text{het,bal}}\left(t_{\text{IR}}\right) \right)\propto \chi^{(2)}(\omega)\text{E}_{\text{IR}}(\omega)\cdot\left[\mathcal{F}\left(\text{E}_{\text{vis}}(t)\text{E}_{\text{LO}}(t)\right)\right]^{*}
\end{equation}
with the Fourier transform operator $\mathcal{F}$ and the complex second order susceptibility $\chi^{(2)}$. The symbol * represents the complex conjugate of the resulting spectrum after Fourier transformation and accounts for the correlation operator in equation \ref{e:het3}. In a last step, all the factors containing the laser fields can be combined to a single complex spectrometer function $\mathcal{S}(\omega)$ simplifying equation \ref{e:fhet1} to
\begin{equation}\label{e:fhet2}
	\mathcal{F}\left(\text{I}_{\text{het,bal}}\left(t_{\text{IR}}\right) \right)\propto \chi^{(2)}(\omega)\cdot\mathcal{S}(\omega)  
\end{equation}
Importantly, the spectrometer function can be obtained by a reference measurement and any additional phase and amplitude effects (\textit{e.g.}\ from linear and nonlinear Fresnel factors \cite{poo11,blo62}) on the involved laser pulses can in principle be included in equation \ref{e:fhet1}. Once the sample independent spectrometer function is determined, phase and amplitude of $\chi^{(2)}$ can precisely be extracted from the interferometric measurement using equation \ref{e:fhet2}.

As shown in Figure \ref{f:tds} we implement this time domain approach in a full collinear beam geometry \cite{xu15,Wang17}. Because the incident LO, IR and vis beams all experience exactly the same optical path extracted phase and amplitude do not depend sensitively on the positioning of the sample and the the measurement is insensitive to vibrations and drifts of optics (including the sample) behind the point of beam combination. Additionally, wave vector conservation now requires that the angle in which the SFG signal is emitted is independent of infrared frequency. This highly simplifies the acquisition of phase resolved SFG spectra in large frequency ranges because there is no need of realignment of the local oscillator and should even allow straightforward extension to ultrabroadband infrared sources \cite{sti17}.

Practically, the time domain approach allows the substitution of single channel detectors (avalanche photo diodes or photomultipliers) for multi-pixel CCD arrays. In general such detectors offer higher sensitivity, lower noise, greater range of wavelength applicability and the possibility of acquiring spectra of each laser shot without loss in sensitivity, all at a fraction of the cost.

\section{Phase and Amplitude Stabilization}
As noted above, our collinear spectrometer is theoretically insensitive to small changes in the common beam path of the three laser pulses. This insensitivity allows us to include an oscillating mirror in the setup that alternately (500 Hz) samples two different spots in the sample area without phase shifts or jitter. By placing the sample and the reference in the spots sampled by the mirror, we are able to perform phase resolved SFG measurements with shot-to-shot referencing (quasi simultaneous referencing). A single experiment thus collects both a sample and a reference spectrum and thus no substituting of a reference in the beam path is necessary (see SI for details of reference measurement calibration).

\begin{figure}[htbp]
	\begin{center}
			\includegraphics[width=0.47\textwidth]{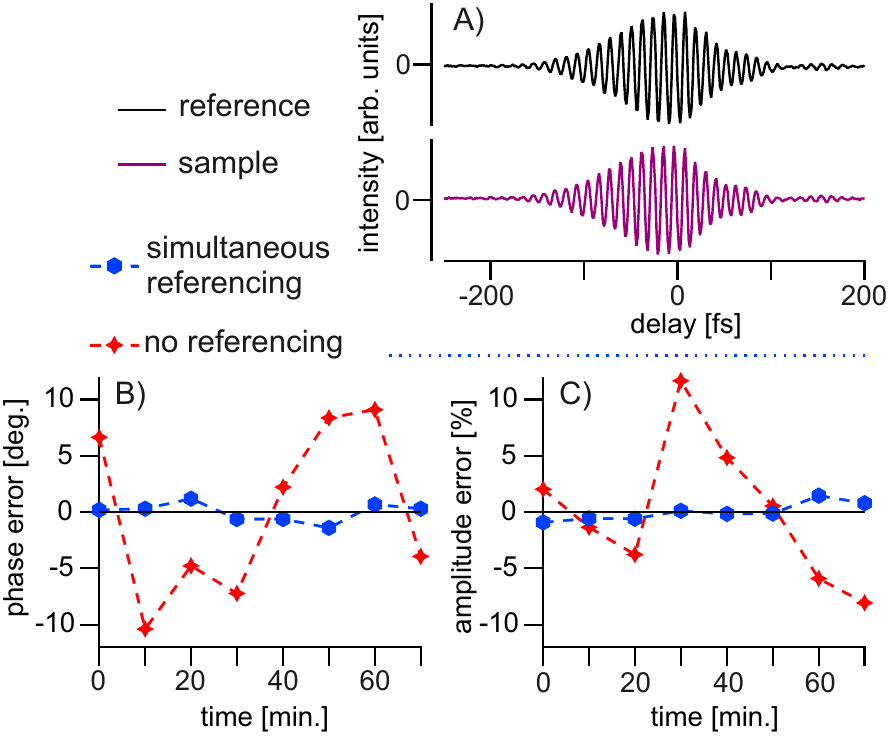}
		\caption{Measurement of the errors in phase and amplitude. Upper panel: A) interferometric raw data, lower panel: evolution of the B) phase and C) amplitude error with time from subsequent interferometric scans. Red crosses: phase/amplitude deviations from the respective mean values (without referencing); blue dots: phase/amplitude deviation from the reference (simultaneous referencing)}
		\label{f:phase_e}
	\end{center}
\end{figure}

We quantified this, theoretical, improvement in phase and amplitude stability by measuring the phase resolved non-resonant SFG response of a spot on a silver mirror, referenced to itself (with the second sampled spot on the same mirror), in consecutive scans over 70 minutes with and without simultaneous referencing. Figure \ref{f:phase_e} shows the resulting phase and amplitude errors for each scan. Clearly without simultaneous referencing, and despite our collinear geometry, significant temporal drifts in both phase ($\approx 20^{\circ}$) and amplitude ($\approx 18\%$) occur over timescales of 1 hour. Performing the measurement with simultaneous referencing (blue dots), these drifts are effectively removed. The residual inaccuracy in phase and amplitude can be estimated by the standard deviations of the measured errors, 0.8 degree and 0.8\% respectively, and are an improvement of one order of magnitude over the non-referenced case.

We note in passing that we found, in preliminary experiments, the relative phase and amplitude between sample and reference to persist for time windows significantly larger than those shown in Figure \ref{f:phase_e} (see \textit{e.g.}\ section 6). Provided collinearity is maintained even spectrometer realignments had little impact. Such stability implies that our ability to conduct very-long heterodyned v-SFG measurements if necessary is limited only by the stability of the laser and the sample. The high precision and stability of the measured phase and amplitude in our v-SFG spectrometer is, to our knowledge, not possible existing heterodyned v-SFG spectrometers.  

\section{Increasing Signal to Noise Ratio}

Given a heterodyned v-SFG spectrometer capable of characterizing phase and amplitude with high precision, our ability to make accurate heterodyned v-SFG measurements is limited by the signal to noise ratio. To evaluate our sensitivity requires quantifying the noise sources. The light intensity generated in our heterodyned SFG experiment can be described by equation \ref{e:het_a} (see supporting information).
\begin{equation}\label{e:het_a}
{{\text{I}}_{het}}({{t}_{\text{IR}}})={{\text{I}}_{\text{LO}}}+\text{I}_{\text{SFG}}^{\max }\cdot \mathcal{A}({{t}_{\text{IR}}})+2\sqrt{{{\text{I}}_{\text{LO}}}I_{\text{SFG}}^{\max }}\cdot \mathcal{J}({{t}_{\text{IR}}})
\end{equation}
I\textsubscript{LO} and I\textsubscript{SFG} are the intensities of the local oscillator and the SFG signal, respectively while the normalized amplitude function $\mathcal{A}({{t}_{IR}})$ accounts for the dependence of the generated SFG signal on t\textsubscript{IR}. The normalized interferogram, $\mathcal{J}({{t}_{IR}})$, contains the desired spectroscopic information. Equation \ref{e:het_a} shows that there are three contributions to the measured intensity:  the LO intensity, which is independent of $t_{\text{IR}}$ and is the first term, the signal from the sample, which slowly varies with $t_{\text{IR}}$ and the third term that describes the interference. The amplitude of the interference signal scales with the square root of the LO intensity and can consequently be increased by raising I\textsubscript{LO}. This suggests that one might use a larger $I_{LO}$ to bring weak signals above the noise floor of the detector \cite{sti08}. However, to evaluate the efficiency of this amplification requires understanding how the overall noise varies as function of the LO intensity.

For reasons of simplicity we consider in what follows the four types of noise that are typically dominant. (i) \emph{Background noise} sources that are independent of light intensity such as dark current from the detector, thermal noise in the electronic components, and readout noise in the acquisition device. Since background noise is independent of the LO intensity signal enhancement directly leads to a reduction of its impact. (ii) \emph{Shot noise} originating in the quantum mechanics of light detection. If the light level at the detector is very low the particle nature of the light becomes apparent in form of shot noise that scales with the square root of the light intensity that reaches the detector $\text{I}_{het}$. With an intense local oscillator $\text{I}_{het}\approx \text{I}_{\text{LO}}$: the shot noise shows the same scaling with the LO intensity as the interference signal. Thus, as pointed out previously by Pool et al \cite{poo11}, the shot noise contribution to the S/N ratio is essentially independent of the LO intensity. (iii) \emph{Signal intensity fluctuations} that originate from fluctuations in laser intensity (see introduction) and scale linearly with signal intensity. For an intense local oscillator the \emph{signal intensity} noise in $\text{I}_{het}$ increases linear with LO intensity. Because the cross term in equation \ref{e:het_a}, the term we wish to isolate, scales with the square root of the LO intensity, increasing I\textsubscript{LO} results in a decrease of the S/N ratio. As a result of this scaling intensity noise quickly grows to the largest noise contribution in heterodyned SFG experiments and is often the main cause of a poor S/N ratio. (iv) \emph{Amplification noise}, originating from the detector and signal processing elements, also generally scales linearly with signal intensity.

\begin{figure}[htbp]
	\begin{center}
			\includegraphics[width=0.45\textwidth]{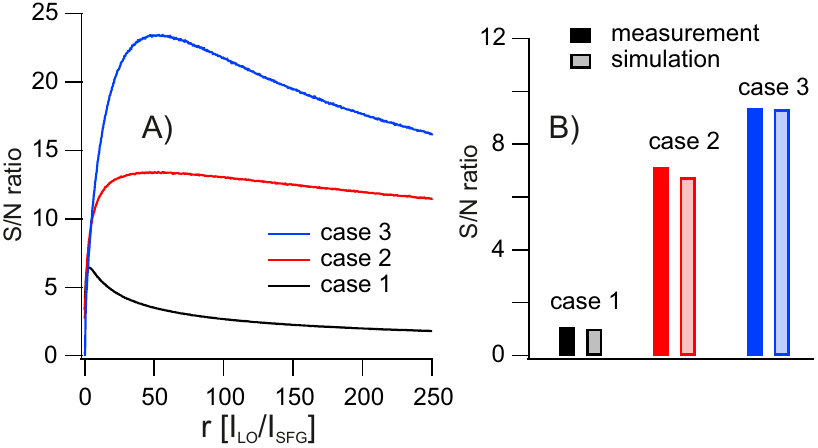}
		\caption{A) Simulation of the signal to noise ratio as function of the local oscillator intensity for a set of test parameters; B) Signal to noise ratios from experiment (solid bars) and simulation (transparent bars); case 1 simple heterodyned SFG; case 2 heterodyned SFG with balanced detection and taking the difference of the detector outputs; case 3 heterodyned SFG with balanced detection and taking the quotient of the difference and the sum of the detector outputs.}
		\label{f:noise_sim}
	\end{center}
\end{figure}

These simple scaling arguments suggest that increasing S/N in $\text{I}_{het}$ requires controlling amplification and signal intensity noise. While the former can only be reduced by a careful choice of ultra-stable detectors and amplifiers, intensity noise can be greatly reduced by employing so-called balanced detection \cite{laa11,ful04,jon09}. In this approach the LO and the sample SFG response are initially set to orthogonal polarization (horizontal and vertical) and thus do not interfere. They are both then propagated through an achromatic waveplate that rotates both polarizations by $45^{\circ}$. All beams are subsequently split again into a horizontal and a vertical polarization component using a polarizing beam splitter. The resulting two beam portions now show interference between the LO and the SFG signals but with opposite signs in the interference term. By simultaneously measuring both intensities in separate detectors (a and b) and subtracting the results one isolates the interference term yielding equation \ref{e:het_d}.
\begin{equation}\label{e:het_d}
{\text{I}}^{(a)}_{het,bal}({{t}_{IR}})-{\text{I}}^{(b)}_{het,bal}({{t}_{IR}})=2\sqrt{{{\text{I}}_{\text{LO}}}\text{I}_{\text{SFG}}^{\max }}\cdot \mathcal{J}({{t}_{IR}})
\end{equation}
The contribution of the \emph{signal intensity noise} now only scales with the square root of the LO intensity reducing its impact on the signal to noise ratio. However, we can go one step further. Intensity noise is present in both beams, the local oscillator and the SFG from the sample. Since both are generated by the same pair of infrared and upconversion pulses their intensity fluctuations are correlated. We can express this mathematically by introducing the noiseless parameter $r$.
\begin{equation}\label{e:het_e}
{{\text{I}}_{\text{LO}}}=r\cdot \text{I}_{\text{SFG}}^{\max }
\end{equation}
Combining equations \ref{e:het_a} and \ref{e:het_e} and forming the quotient between the difference and the sum of the detector outputs yields
\begin{equation}\label{e:het_h}
\frac{{\text{I}}^{(a)}_{het,bal}({{t}_{\text{IR}}})-{\text{I}}^{(b)}_{het,bal}({{t}_{\text{IR}}})}{{\text{I}}^{(a)}_{het,bal}({{t}_{\text{IR}}})+{\text{I}}^{(b)}_{het,bal}({{t}_{\text{IR}}})}\approx \frac{2\cdot \mathcal{J}({{t}_{\text{IR}}})}{\sqrt{r}}
\end{equation}
The quotient in equation \ref{e:het_h} is now a quantity that is free from any intensity noise. However, this is only strictly valid in absence of other noise contributions. To evaluate the improvement in S/N under realistic conditions we performed a noise simulation described in detail in the supporting information. 
In the simulation we determined the S/N ratio (with a set of physically plausible values for the four noise contributions) as function of the LO intensity for three cases:  simple heterodyning (based on equation \ref{e:het_a}), heterodyning with balanced detection and taking the difference of the detector outputs (equation \ref{e:het_d}), and heterodyning with balanced detection taking the quotient of the difference and the sum (equation \ref{e:het_h}). The result of the simulation is depicted in figure \ref{f:noise_sim}A) showing a clear improvement in S/N ratio for the cases 2 and 3. As we show in detail in the SI, the amount of improvement is a function of the exact composition of the overall noise. As expected, the improvement increases with growing relative contribution of the intensity noise. Another important result from the simulation is the existence of a maximum in the signal to noise ratio at a particular intensity of the local oscillator (in the example at $r\approx 50)$. The exact position of this maximum depends again on the details of the noise composition and must be determined based on the noise characteristics of the spectrometer in use. The intensity of the LO can then be tuned to this value to maximize the S/N of the spectrometer.
\begin{figure*}
	\begin{center}
			\includegraphics[width=0.9\textwidth]{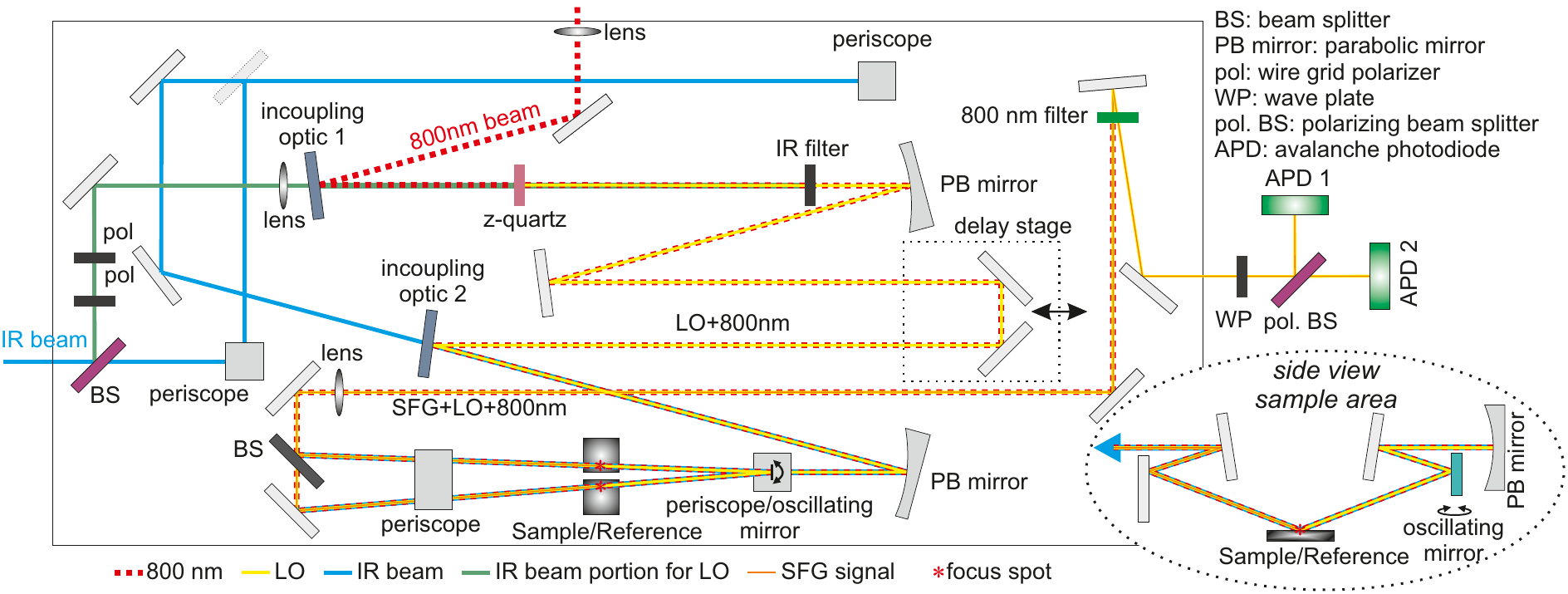}
		\caption{Diagram of the interferometer. The main portion shows the entire interferometer in the top view. The inset shows the sample area viewed from the side.}
		\label{f:interfer}
	\end{center}
\end{figure*}

To test the improvement of the S/N ratio in our spectrometer experimentally the time delay t\textsubscript{IR} was moved to the maximum in the interferogram and the heterodyned SFG signals from a gold surface were recorded for 10,000 laser shots. From these data traces the S/N ratios were extracted according to cases 1 to 3. The results are presented in figure \ref{f:noise_sim}B) (solid bars). Again, we see a significant enhancement of the S/N ratio for cases 2 and 3. The total improvement amounts to one order of magnitude compared to the simple heterodyned case (case 3 vs. case 1). Note that this improvement corresponds to a decrease in the required averaging by about two orders of magnitude. The quantitative agreement between simulation and experiment clearly shows both that the four types of noise we identify are sufficient to describe our spectrometer's performance and that balanced detection effectively eliminates the largest contributor \footnote{For details of the experimental quantification of each noise source, by blocking different beams in the spectrometer, see the Supporting Information}. 
 
\section{Experimental}
Infrared and 800 nm upconversion pulses are generated by a commercial, Ti:Sapphire based,  1 kHz, regenerative amplifier that produces femtosecond pulses at 800 nm. A portion of this pulse energy is then converted to the infrared using a commercially available optical parametric amplifier and difference frequency generation set-up (for details see Supporting Information) and sent into the interferometer. The design of the interferometer is depicted in Figure \ref{f:interfer}. The infrared beam that enters the interferometer is split into two portions by a beam splitter (ZnSe window). The weak (ca. 8\%) reflected portion is used for generation of the local oscillator and the strong transmitted for sum frequency generation at the sample. The reflected part passes through two free standing wire grid polarizers (Infraspecs) allowing for tunable attenuation and polarization control. The beam is then transmitted through the first incoupling optic, \textit{i.e.}\ a 2 mm thick Ge window with a custom coating that is highly reflective in the visible to near infrared (upconversion beam and LO) and highly transmissive in the mid-infrared (2.5-13 $\mu$m). At the surface of this incoupling optic the 800 nm upconversion beam is superimposed onto the infrared in a collinear fashion. Two lenses (20 cm \ce{CaF2}, and 100 cm BK7 for the infrared and the visible beams, respectively) that are placed in each beam path before the incoupling optic focus the two beams into a common spot in a thin z-cut quartz window (50 $\mu$m thick) generating a weak sum frequency signal (the local oscillator, LO). 

The intensity and polarization of the local oscillator (LO) can be controlled by adjusting the polarization and intensity of the infrared pulses using the two polarizers inside the IR beam path in combination with rotating the quartz wafer about its z-axis. This is possible due to the symmetry of the second order susceptibility tensor of $\alpha$-quartz \cite{lu03}. One therefore maintains full tunability of the LO at any given intensity and polarization of the upconversion beam. Behind the quartz wafer three beams are co-propagating colinearly, the 800 nm, the LO, and the infrared. The infrared is filtered out by a thin shortpass filter, while the remaining (800 nm and LO) are collimated by an off axis (15$^{\circ}$) parabolic mirror. The two beams subsequently enter a delay line with a computer controlled miniature piezo translation stage (PI, Q-521-330) before they get combined with the second infrared beam at the surface of a second incoupling optic (same type as incoupling optic 1). At this point, there are again three laser beams co-propagating , the 800 nm upconversion beam, the local oscillator and the second infrared portion. The three pulses are then focused by a second off axis (15 deg.) parabolic mirror. Using a two-mirror periscope right behind the parabolic mirror (see inset Figure \ref{f:interfer}) the beams are directed downwards at an incidence angle of 72 degree onto the sample which is placed horizontally in the focus of the three beams. The second mirror in this periscope is an oscillating mirror mounted on a galvo motor that oscillates at 500 Hz alternately sampling two different spots at the sample position (separation of ca. 1 cm).

After the sample the reflected beams pass through a second mirror periscope before the two different beam paths (sample and reference) are recombined on a beam splitter. The generated SFG signals (LO and SFG from the sample/reference) are collimated by a lens and the 800 nm and IR frequencies are filtered out by two stacked shortpass filters. The laser beam now consists of the pure heterodyned SFG signal from the sample/reference, leaves the interferometer and is detected employing balanced detection. After passing through an achromatic waveplate the beams are split into two portions by a polarizing beam splitter cube (Thorlabs, CCM1-PBS252). The two resulting beams are each focused onto an avalanche photodiode (Thorlabs, APD410A2) where the intensity of the heterodyned SFG signal is measured. The signals from the two APDs are then integrated by gated integrators (SRS, Boxcar Averager) and finally digitalized. More experimental details including data acquisition and treatment as well as the calibration of the translation stage movement are given in the supporting information.

\section{Example: SFG Experiment on $\alpha$-Quartz}
\begin{figure*}
	\begin{center}
			\includegraphics[width=0.9\textwidth]{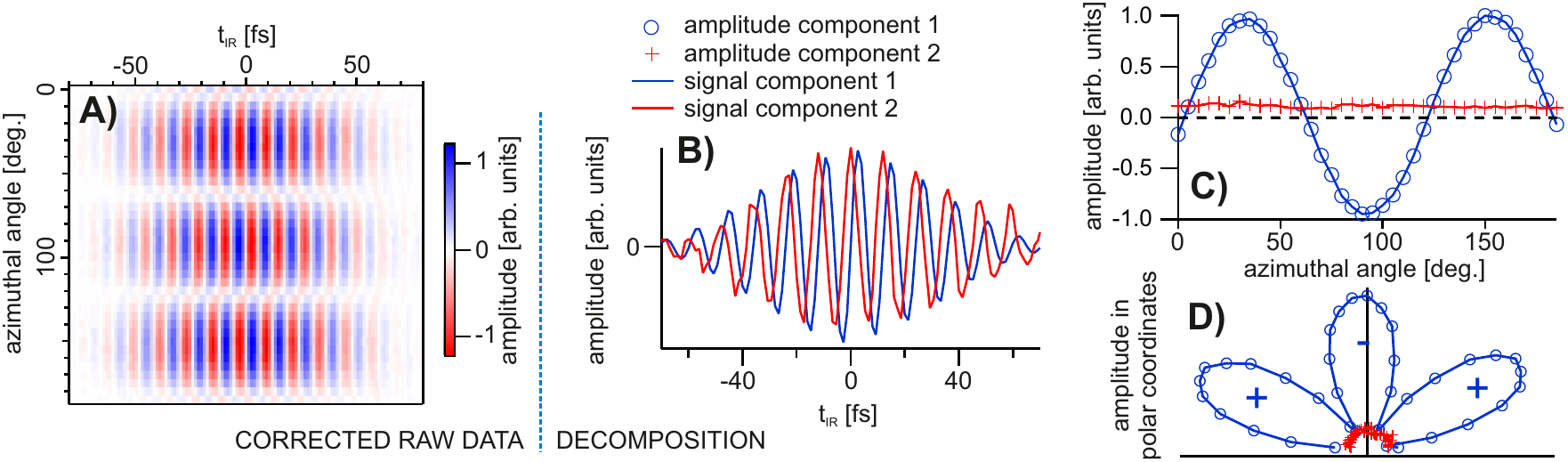}
		\caption{Referenced, phase resolved SFG measurement of the z-cut quartz surface. A) Interferometric raw data as function of the azimuthal angle (corrected for drifts in phase and amplitude); B) time domain signals from the two extracted components; C) azimuthal dependence of the amplitude for the two components; D) same as C) represented in polar coordinates.}
		\label{f:quartz}
	\end{center}
\end{figure*}
With this new spectrometer in hand we demonstrate its performance by showing first results that we obtained from phase resolved SFG measurements of $\alpha$-quartz. Due to its non-centrosymmetric crystal structure $\alpha$-quartz is bulk SFG active. The off resonant bulk SFG response has been well studied and is commonly used as internal or external phase reference \cite{lu03,Ohno16}. However, a phase resolved isolation of the SFG contribution from the surface has so far not been achieved. The effective second order susceptibility $\chi_{_{eff}}^{(2)}$ that is measured in SFG experiments performed in external reflection geometry can be expressed by the following equation
\begin{equation}\label{e:b_s}
{\chi _{_{eff}}^{(2)}=\chi _{S}^{(2)}+i\chi _{B}^{(2)}}
\end{equation}
where $\chi _{S}^{(2)}$ and $\chi _{B}^{(2)}$ represent the surface and bulk contributions, respectively. If all interacting laser fields are off resonant $\chi _{S}^{(2)}$ and $\chi _{B}^{(2)}$ are real but their contributions to the effective susceptibility are phase shifted by 90 deg. with respect to each other. The phase of the optical response is therefore a good indicator to determine its origin. Several studies have shown that $\chi_{_{eff}}^{(2)}$ in the case of $\alpha$-quartz is dominated by the bulk response, the measured susceptibility is therefore in good approximation imaginary \cite{sun16}. The bulk nonlinear susceptibility has the well-known threefold azimuthal symmetry which is governed by the crystal symmetry. The surface contribution, however, should have at least one component that originates from the macroscopic potential asymmetry in the direction along the surface normal. This component is isotropic in the surface plane and should therefore possess ${{C}_{\infty }}$ symmetry. The isolation of the isotropic surface signal is, however, far from trivial: Compared to $\chi _{B}^{(2)}$ the isotropic component is small and is therefore completely buried under the bulk response. Furthermore, the two contributions only differ in their phases and symmetry. On the other hand, this challenging task is precisely the type of problem that we wish to solve with our spectrometer. 

We therefore measured the off resonant SFG response from a z-cut quartz sample in ppp \footnote{ppp means that the infrared, the upconversion, and the detected SFG pulses are p polarized} polarization as function of the azimuthal angle with our phase resolved spectrometer. A second, stationary z-cut quartz was placed in the second sampled spot serving as reference. To ensure pure off resonant interactions the experiment was carried out at an infrared frequency of 2800 $c{{m}^{-1}}$. For each azimuthal angle (scanned in steps of 5 deg.) the entire interferogram from sample and reference was recorded. In a subsequent step the resulting sample interferograms were corrected for changes in spectral phase and amplitude in the corresponding reference spectra (removing any temporal drifts in phase and amplitude over the course of the experiment). The corrected interferometric raw data are depicted in figure \ref{f:quartz}A). 

At a first glance the result shows the well-known threefold symmetry of the bulk contribution, however, a closer look into the regions where the bulk contribution vanishes reveals that the phases in the interferograms shift. This indicates the presence of at least a second contribution. To isolate this second component we performed a linear decomposition of the time-domain data based on singular value decomposition and found that the data can indeed be well described by the superposition of two components with different symmetry. The azimuthal dependence of these two components and the corresponding time-domain signals are shown in figures \ref{f:quartz}B) and C). The first component shows the expected threefold symmetry whereas the second component is isotropic. Furthermore, comparing the corresponding time domain signals shows a relative phase shift of 90 deg. between the two signals while their magnitude spectra are nearly equal. This suggests the isotropic contribution originates from the sample surface. The overall amplitude of this second component is meanwhile about ten times smaller than the bulk contribution (but still well resolved). This corresponds to a difference in intensity of the two radiated signals of two orders of magnitude which explains why it is so difficulty to detect this signal with homodyned SFG spectrometers.

The perfect match between these experimental results and the theoretical considerations shown above are in strong favor of our preliminary interpretation that the isolated second component is indeed the isotropic surface contribution of the measured second order susceptibility. To our knowledge, this signal has not been directly measured before. For a detailed characterization of this second contribution more experiments and analysis are obviously required, however, this is beyond the scope of this article. What this experiment demonstrates, however, is that we are now indeed capable of retrieving a small SFG signal that is buried in a complex SFG spectrum. The decomposition does not only reveal its contribution but also recovers its complex spectrum. As shown before, the analysis of the phases even allows us to attribute the different signals to surface and bulk origin, respectively. The success of the decomposition clearly shows that over the entire duration of the experiment (4 h) even smallest drifts in the spectral phases and amplitudes were successfully suppressed. 

\section{Summary and Outlook}

We have described a new spectrometer design that allows for the acquisition of extended data sets of complex, low noise SFG spectra with high accuracy in spectral phase and amplitude. The high precision of the data makes it possible to use linear numerical algebra methods (\textit{e.g.}\ singular value decomposition) for the decomposition of the spectra into their different contributions as shown the section 6. With these possibilities in hand we can extent our SFG studies towards species whose cross section is too weak or whose interface population is too small to be detected with the current state-of-the-art instruments. A possible application is \textit{e.g.}\ the investigation of the structure and dynamics of protons at interfaces, a topic which is currently heavily debated \cite{Noam16} and a system where the precision that we achieved with our technique will be crucial. 
%%%END OF MAIN TEXT%%%

%\section*{Conflicts of interest}
%There are no conflicts to declare.


%\balance


%%%REFERENCES%%%
%\bibliography{setup_paper_bib} %You need to replace "rsc" on this line with the name of your .bib file
%\bibliographystyle{rsc} %the RSC's .bst file

\providecommand{\latin}[1]{#1}
\makeatletter
\providecommand{\doi}
  {\begingroup\let\do\@makeother\dospecials
  \catcode`\{=1 \catcode`\}=2\doi@aux}
\providecommand{\doi@aux}[1]{\endgroup\texttt{#1}}
\makeatother
\providecommand*\mcitethebibliography{\thebibliography}
\csname @ifundefined\endcsname{endmcitethebibliography}
  {\let\endmcitethebibliography\endthebibliography}{}

\end{document}